\documentclass[a4paper,12pt]{article}
\usepackage{epsfig}
\usepackage{graphicx}
\usepackage{subfigure}
\begin{document}

\title{\textbf{Collective Coordinate Approach to the Dynamics of Various Soliton-Obstruction Systems}}
\author{Jassem H. Al-Alawi\thanks{e-mail address:J.H.Al-Alawi@durham.ac.uk}
\\ Department of Mathematical Sciences,University of Durham, \\
 Durham DH1 3LE, UK\\
}

\date{\today}

\maketitle

\begin{abstract}
Various soliton-obstruction systems have been studied from analytical perspective. We have used collective coordinate to approach the dynamics of solitons as they meet a potential obstruction in a form of square barriers and holes for three models in (1+1) dimensions, namely: $\lambda\phi^{4}$ model, deformed Sine-Gordon model, and a model that give rise to Q-ball solution. We have shown that our approximated field solution is valid enough to describe the behaviour of solitons scattering off a potential obstruction.
\end{abstract}

\section{\textbf{Introduction}}
Solitons scattering from obstructions have been studied numerically ({\it eg} \cite{PZ05, PZ07,KJ06, AZ07, AZ08, AZ09}) and have shown interesting dynamics. As we have seen in our previous study of soliton scattering that each model has its on dynamical features. However, all models have as observed numerically mutual dynamics. In all models investigated so far solitons have elastic scattering on barriers and inelastic scattering on holes. Furthermore, the scattering on barriers have revealed that the core of solitons is not important and the dynamics in this case resembles the scattering of a featureless point particle. Therefore, results obtained from the dynamics on barriers can be worked out analytically. On the contrary, the scattering on holes have shown some challenging dynamics. Solitons in all models investigated loses some their energy which makes the analytical description of such behaviour more complicated. Mo, solitons have shown quantum -like behaviour as observed in many studies and this makes the behaviour quite hard to be understood  within a classical theory.

  In this paper, we will try to shed some light on the dynamics of various soliton-obstruction systems from analytical perspective and compare the analytical results with the ones observed in numerical simulations. To approach the dynamics of the soliton-obstruction systems we will use collective coordinates, {\it ie} the parameters of the field solution. We will, for simplicity,  approximate the solitonic solution by only one parameter which represents the position of soliton. Hence, the soliton solution can be approximated as $\phi\left(\,x,\,t\right)\approx\phi\left(\,x-\,X\left(\,t\right)\right)$, where $\,X\left(t\right)$ is the position of soliton. 

Soliton field has an infinite degrees of freedom and so a more reliable approximation would be if we allow the kink of the soilton to vary at the obstruction, {\it ie} $\varphi\left(\,x,\,t\right)\sim\varphi\left(\beta\left(\,t\right)\left(\,x-\,X\left(t\right)\right)\right)$ where $\beta\left(\,t\right)$ is the parameter that describes the variation of the kink at the obstruction. So, Far away from the obstruction at $\,t\rightarrow\pm\infty$, $\beta\rightarrow1$. But, the resulting dynamics is more complicated to handle. Thus, we will consider only an approximated field with one parameter that describes the position of the soliton. 

 In following sections we will study three different soliton-obstruction systems in (1+1) dimensions and show how much our approximations are valid. 

\section{\textbf{The Central Idea of Approaching the Dynamics of Soliton-Obstruction Systems}}

We will Consider Lagrangian densities for which a soliton solution can be obtained. A Lagrangian density that describes soliton-obstruction systems is given by 

\begin{equation}
 \mathcal{L}=\,T-\,\tilde V,
\end{equation}

where $\,T$ and $\,\tilde V$ are the kinetic and the potential terms respectively. The obstruction is introduced to the Lagrangian density via a coupling in the potential term. The obstruction is confined in a certain region of space and it is seen by a soliton as an external potential hole or a barrier.

where $\,\tilde V=\tilde\lambda\left(\,x\right)\,V$ and $\tilde\lambda=1+\lambda\left(\,x\right)$. The obstruction is localised in a finite region of space and are either square wells or square barriers. So, in what follows, we will write $\lambda\left(\,x\right)=\lambda_{0}$. Thus we can describe them using Heaviside functions which are defined by:

\begin{equation}
H\left(x\right)=\cases{ 0 & \,x \textless0 \cr   1 & \,x \textgreater0 \cr}.
\end{equation}

Therefore the potential can be written as:

\begin{equation}
 \tilde\lambda\left(\,x\right)=1+\lambda_{0}\left[\,H\left(\,x+\,x_{0}\right)-\,H\left(\,x-\,x_{0}\right)\right],
\end{equation}

where $\,x_{0}$ is the position of the obstruction and $\lambda_{0}$ is the parameter that describes the magnitude ({\it ie} height or depth) of the obstruction. A study that has been conducted in this regard for Sine-Gordon model  in \cite{KJ08}, has considered the obstruction as a one point perturbation. In our study we will over look the dynamics that span the space for which the obstruction is localised. The sign of $\lambda_{0}$ determines whether the potential obstruction is a hole or a barrier. When $\lambda_{0}>0$, the obstruction is a barrier and a hole when $\lambda_{0}<0$.
In the following sections we will analyse various soliton-obstruction systems that would shed a light on the dynamics of such systems. So, we need to calculate the Lagrangian which is given by

\begin{eqnarray}
&&\,L=\int^{\infty}_{-\infty}dx\mathcal{L}\nonumber\\
&&=\int^{\infty}_{-\infty}dx\left(T-\left(1+\lambda_{0}\left[H\left(x+x_{0}\right)-H\left(x-x_{0}\right)\right]\right)\,V\left(x\right)\right)\nonumber\\
&&=\int^{\infty}_{-\infty}dx\left(T-V\right)-\lambda_{0}\int^{x_{0}}_{-x_{0}}dxV\left(x\right).
\end{eqnarray}

In the following sections we will build an approximate field solutions for various soliton-obstruction systems and give analytical description for the dynamics of these models and compare our analytical description with the numerical simulation work that have been explained in the previous papers \cite{AZ07,AZ08,AZ09}.

\section{\textbf{$\lambda\phi^{4}$ Model}}

The Lagrangian density for the $\lambda\phi^{4}$ model is given by

\begin{equation}
 \mathcal{L}=\frac{1}{2}\dot\varphi^{2}-\frac{1}{2}\varphi^{\prime^{2}}-\tilde\lambda\left(\varphi^{2}-1\right)^{2},
\end{equation}

We use the ansatz

\begin{equation}
 \varphi\left(\,x;\,X\right)=\,tanh\left[\sqrt{2}\left(\,x-\,X\left(\,t\right)\right)\right],
\end{equation}

where $\,X\left(\,t\right)$ is the position of soliton as a function of time.

Now, the Lagrangian density, after substituting the above results, becomes
\begin{eqnarray}
 &&\mathcal{\,L}=\left(\,\dot X^{2}-2\right)\,sech^{4}\left[\sqrt{2}\left(\,x-\,X\right)\right]\nonumber\\
&&\quad\quad\quad\quad-\lambda_{0}\left[\,H\left(\,x+\,x_{0}\right)-\,H\left(\,x-\,x_{0}\right)\right]\,sech^{4}\left[\sqrt{2}\left(\,x-\,X\right)\right].
\end{eqnarray}

The Lagrangian:

\begin{eqnarray}
&&\,L=\int^{\infty}_{-\infty}\mathcal{L}\,dx\nonumber\\
&&=\left(\,\dot X^{2}-2\right)\frac{4}{3\sqrt{2}}-\lambda_{0}\int^{\,x_{0}}_{-\,x_{0}}\,sech^{4}\left[\sqrt{2}\left(\,x-\,X\right)\right] \nonumber\\
&&=\left(\,\dot X^{2}-2\right)\frac{4}{3\sqrt{2}} \nonumber\\
&&-\frac{\sqrt{2}}{6}\lambda_{0}\,tanh\left(\sqrt{2}\left(\,X+\,x_{0}\right)\right)\left[\,sech^{2}\left(\sqrt{2}\left(\,X+\,x_{0}\right)\right)+2\right]\nonumber\\
&&+\frac{\sqrt{2}}{6}\lambda_{0}\,tanh\left(\sqrt{2}\left(\,X-\,x_{0}\right)\right)\left[\,sech^{2}\left(\sqrt{2}\left(\,X-\,x_{0}\right)\right)+2\right] 
\end{eqnarray}

The potential is given by

\begin{equation}
 \,V\left(\,X\right)=\frac{\sqrt{2}}{6}\lambda_{0}\left[\begin{array}{cc}&\,tanh\left[\sqrt{2}\left(\,X+\,x_{0}\right)\right]\left(\,sech^{2}\left[\sqrt{2}\left(\,X+\,x_{0}\right)\right]+2\right)\\

&-\,tanh\left[\sqrt{2}\left(\,X-\,x_{0}\right)\right]\left(\,sech^{2}\left[\sqrt{2}\left(\,X-\,x_{0}\right)\right]+2\right) \end{array}\right]
\end{equation}

Fig. 1  and fig. 2  show the potential as a function of the position of soliton for $\lambda_{0}=\pm1$ respectively when the obstruction is located at $\vert\,x_{0}\vert\leq5$.

\begin{figure}
\begin{center}
\includegraphics[angle=270, width=10cm]{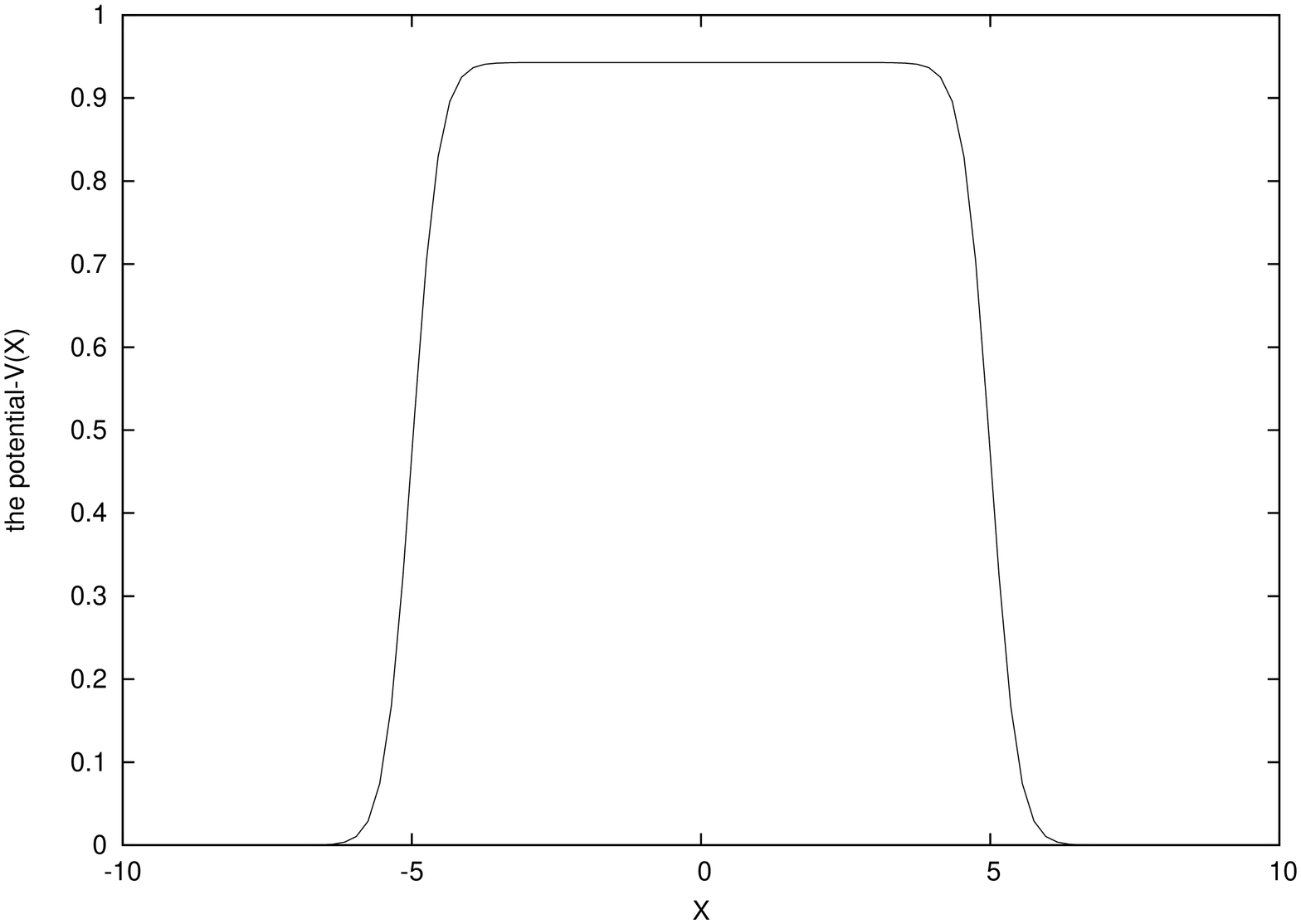}
\caption{The potential raised by a barrier, $\lambda_{0}$=1}
\end{center}
\end{figure}

\begin{figure}
\begin{center}
\includegraphics[angle=270, width=10cm]{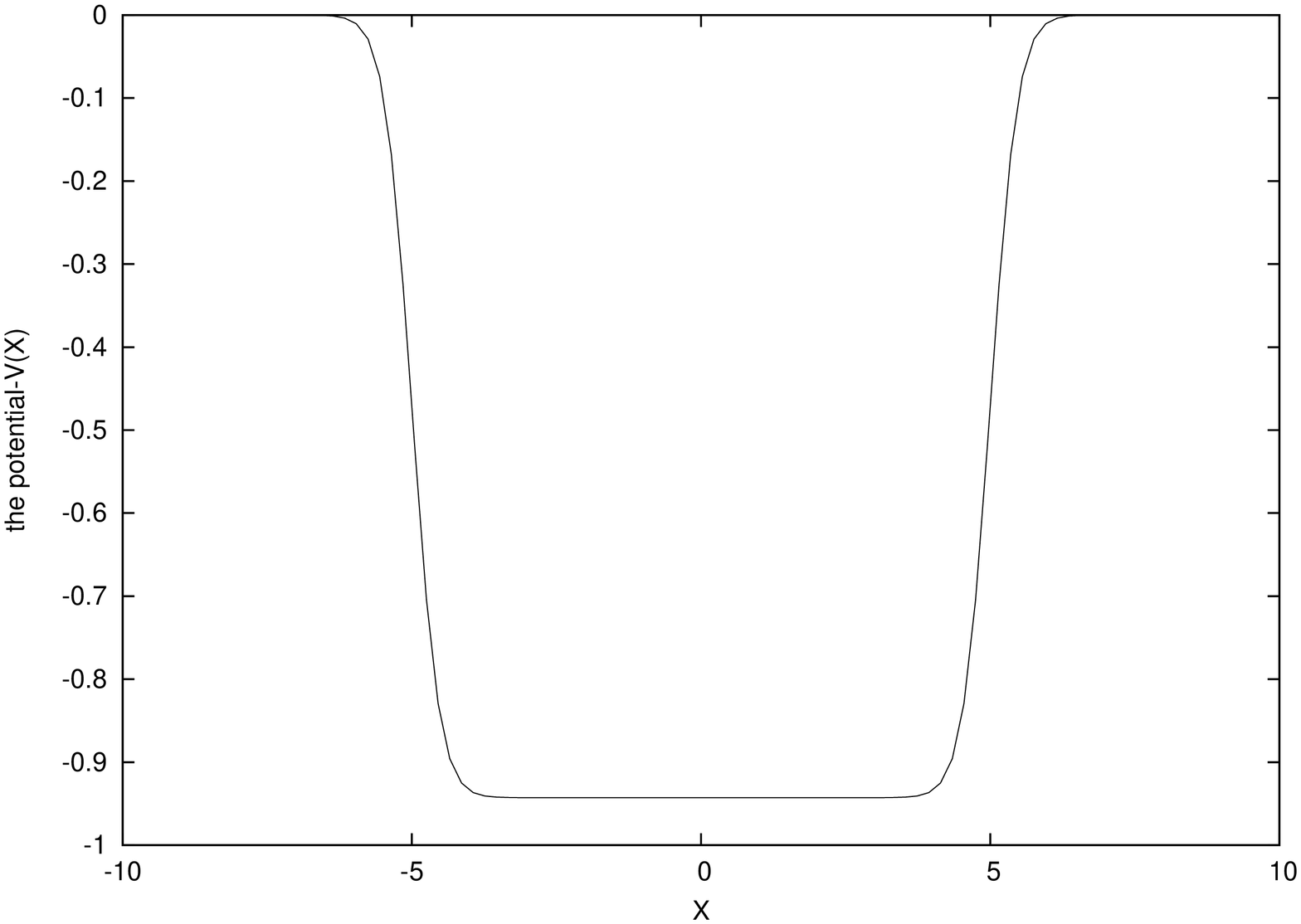}
\caption{The potential raised by a hole, $\lambda_{0}$=1}
\end{center}
\end{figure}

The potential exponentially suppressed outside the obstruction's region. 

Using the Lagrange-Euler equation,$\frac{\partial\,L}{\partial\,X}-\frac{\,d}{\,dt}\left(\frac{\partial\,L}{\partial\,\dot X}\right)=0$, the e.o.m is 

\begin{equation} \label{eommod1}
 \frac{8}{3\sqrt{2}}\,\ddot X+\lambda_{0}\left[,sech^{4}\left(\sqrt{2}\left(\,X+\,x_{0}\right)\right)-\,sech^{4}\left(\sqrt{2}\left(\,X-\,x_{0}\right)\right)\right]=0.
\end{equation}

The force of the obstruction on soliton is $\,F=\,M\,\ddot X$ ($\,M$ is the rest mass of soliton) and is given by

\begin{equation}
 \,F=-\lambda_{0}\left[\,sech^{4}\left(\sqrt{2}\left(\,X+\,x_{0}\right)\right)-\,sech^{4}\left(\sqrt{2}\left(\,X-\,x_{0}\right)\right)\right],
\end{equation}

where the rest mass $\,M_{rest}=\frac{8}{3\sqrt{2}}$.

In case of a barrier, $\lambda_{0}>0$,  then the force is repulsive as $\,F<0$. The case of a hole where $\lambda_{0}<0$ the force is attractive since $\,F>0$. Fig 3  and fig. 4 show the force exerted by the obstruction on a soliton for $\lambda_{0}=\pm1$ when the obstruction is inserted in the space region, $\vert\,x_{0}\vert\leq5$. This is in agreement with the observed behaviour. In case of a barrier the force for the first half of the barrier is repulsive but it becomes attractive for the second half. In the hole it is the other way around.

\begin{figure}
\begin{center}
\includegraphics[angle=270, width=8cm]{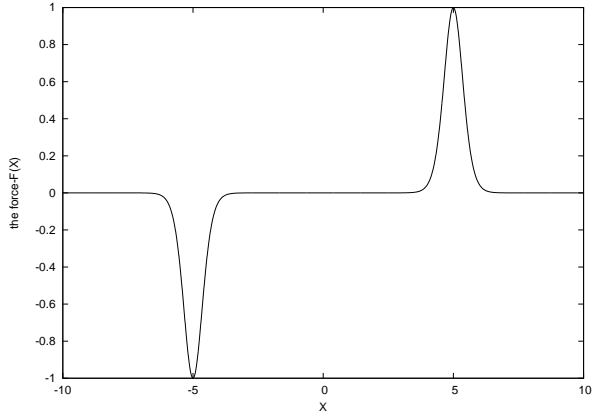}
\caption{The force on the soliton by a barrier, $\lambda_{0}=1$}
\end{center}
\end{figure}

\begin{figure}
\begin{center}
\includegraphics[angle=270, width=8cm]{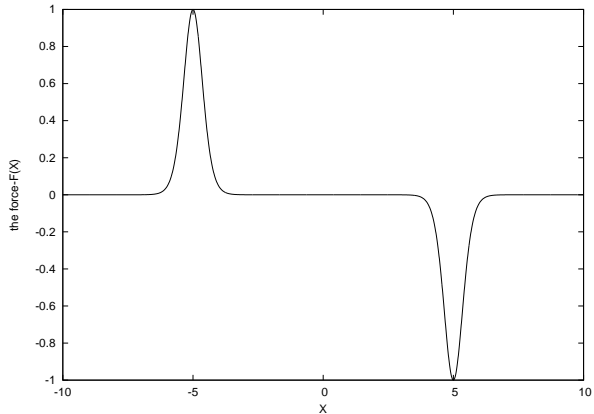}
\caption{The force on the soliton by a hole, $\lambda_{0}=-1$}
\end{center}
\end{figure}

The force as can be seen in fig. 3 and fig. 4  decays out exponentially as we move away from the obstruction and increases, in magnitude, exponentially at the edges of the obstruction. Then it dies out exponentially when the soliton is inside the hole or at the top of the barrier. Thus, when the soliton is far away from obstruction, {\it $\,X\rightarrow\pm\infty$}, $\lambda_{0}=0$, the force is zero and so

\begin{equation}
 \,\ddot X=0.
\end{equation}

 The solution is 

\begin{equation}
 \,X=\,X_{0}+\,u\,t,
\end{equation}

where $\,X_{0} $ is the initial position of soliton.

We have solved the equation of motion (\ref{eommod1}) with initial conditions that specify the values of the position $\,X\left(0\right)$
of the soliton and its speed using fourth order Runge Kutta method. Fig. 5 and fig. 6 show the trajectories of the soliton starting from an initial position,   $\,X\left(0\right)=-15$. These trajectories describe a soliton moving with a speed of 0.5 and interacting with a barrier of height $\lambda_{0}$=0.25 and a hole of depth -0.25. The figures demonstrate the validity of our approximation. We have found such agreement between our analytical approach and the numerical simulations in many occasions.

\begin{figure}
\begin{center}
\includegraphics[angle=270, width=8cm]{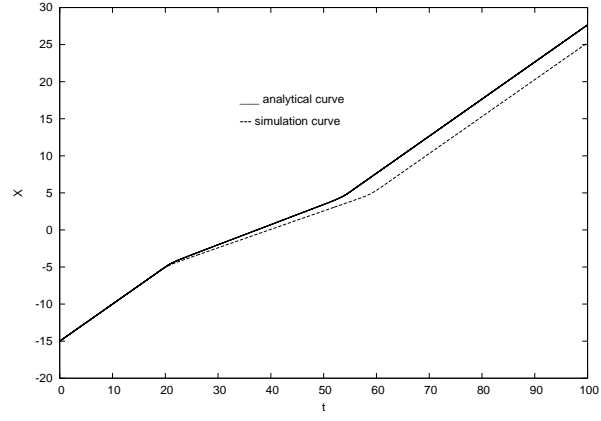}
\caption{The trajectories of soliton-barrier system}
\end{center}
\end{figure}

\begin{figure}
\begin{center}
\includegraphics[angle=270, width=8cm]{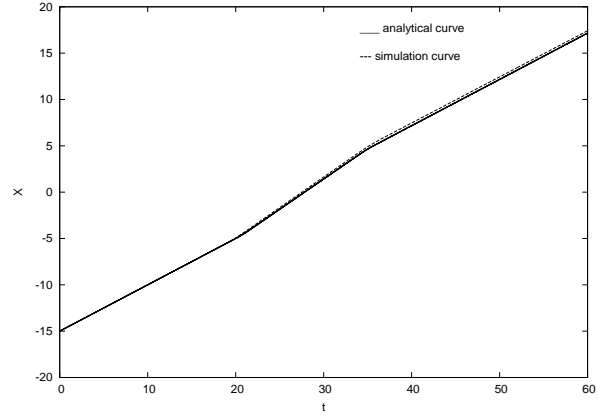}
\caption{The trajectories of soliton-hole system}
\end{center}
\end{figure}

We can calculate the total energy of soliton.

\begin{eqnarray}
&&\,E = \frac{\partial\,L}{\partial\,\dot X}\,\dot X-\,L \nonumber\\
 && = \frac{4}{3\sqrt{2}}\left(\,\dot X^{2}+2\right)+\frac{\sqrt{2}}{6}\lambda_{0}\left[\begin{array}{cc} &\,tanh\left(\sqrt{2}\left(\,X+\,x_{0}\right)\right)\left[\,sech^{2}\left(\sqrt{2}\left(\,X+\,x_{0}\right)\right)+2\right]\\

&-\,tanh\left[\sqrt{2}\left(\,X-\,x_{0}\right)\right]\left[\,sech^{2}\left(\sqrt{2}\left(\,X-\,x_{0}\right)\right)+2\right]\end{array}\right].
\end{eqnarray}

The total energy, when the soliton is away from the potential, $\lambda_{0}=0$, is 

\begin{equation}
 \,E=\frac{1}{2}\left(\frac{8}{3\sqrt{2}}\right)\,\dot X^{2}+\frac{8}{3\sqrt{2}}.
\end{equation}

And so when the soliton is at rest,{\it i.e} $\,\dot X=0$, the total energy is the mass of soliton:

\begin{equation}
 \,M_{rest}=\frac{8}{3\sqrt{2}}.
\end{equation}

This is, in fact, the minimum energy of the soliton {\it i.e} $\,E_{min}=\,M_{rest}$.

Therefore, 

\begin{equation}
 \,E\geq\frac{8}{3\sqrt{2}}.
\end{equation}

In order to find mass of soliton at the top of a barrier or inside a hole with $\,\dot X=0$ we need to position the soliton where the force of the barrier or a hole on the soliton is minimum. The best position is when $\,X=0$ since the force is zero. However, one needs to keep in mind that a soliton is not a point particle but an extended structure where there will be always an interaction between the tail of the soliton and the tail of the force on both ends.  At $\,X=0$, the total energy which is the mass of a soliton at the obstruction is given by 

\begin{eqnarray}
&&\,M=\,M_{rest}+\frac{\sqrt{2}}{3}\lambda_{0}\,tanh\left(\sqrt{2}\,x_{0}\right)\left(\,sech^{2}\left(\sqrt{2}\,x_{0}\right)+2\right)\nonumber\\
&&=\,M_{rest}+0.943\lambda_{0}.
\end{eqnarray}

Table 1 and table 2 show the numerically and calculated  masses of soliton at various $\lambda_{0}$ when $\,x_{0}=5$
for various barrier heights and hole depths.

\begin{table}
\begin{center}
\begin{tabular}{ccc} \\[3pt]
\hline
$\lambda_{0}$ & calculated $\,M_{B}$ & observed $\,M_{B}$ \\ [3pt]
\hline
 0.25 &2.12131 &  2.12126 \\
 0.5 & 2.35701 &  2.35696  \\
 0.75  & 2.59272 &   2.59262 \\ [3pt]
\hline
\end{tabular}
\end{center}
\caption{The calculated and observed masses of a soliton at the top of barriers, $\,M_{B}$}
\end{table}

\begin{table}
 \begin{center}
\begin{tabular}{ccc}\\ [3pt]
\hline
$\lambda_{0}$ & calculated $\,M_{H}$ & observed $\,M_{H}$   \\ [3pt]
\hline
 -0.25 &1.6499 &  1.6499 \\
 -0.5 &1.4142  &  1.4142  \\
 -0.75  & 1.1785 &  1.1785  \\ [3pt]
\hline
\end{tabular}
\end{center}
\caption{The calculated and observed masses of a soliton inside holes, $\,M_{H}$}
\end{table}

Thus, the tables1 and 2 confirm the excellent agreement between the analytical and numerical values.

The total energy must be conserved. So, the energy away from the obstruction must equal to the energy during which the soliton interacting with the obstruction. Thus, if a soliton is moving with a velocity $\,u$ and experiences an obstruction, {\it ie} a barrier, then energy conservation implies 

\begin{equation}
 \frac{1}{\sqrt{1-\,u^{2}}}\,M_{rest}=\frac{1}{\sqrt{1-\,u_{b}^{2}}}\,M,
\end{equation}

where $\,u_{b}$ is the velocity of the soliton at the obstruction and $\,M$ is the mass of the soliton at the top of a barrier or inside a hole. Hence, we can  with a good approximation calculate the velocity of soliton when it is crossing the region of the obstruction

\begin{equation}
 \,u_{b}=\sqrt{1-\left(\frac{\,M}{\,M_{rest}}\right)^{2}\left(1-\,u^{2}\right)}
\end{equation}


To calculate the critical velocity,$\,u_{c}$, in case of a barrier we set $\,u=0$ and the equation reduces to

\begin{equation}\label{ucrmod1}
 \,u_{c}=\sqrt{1-\left(\frac{\,M_{B}}{\,M_{rest}}\right)^{2}}
\end{equation}






 



 When the soliton moves with the critical velocity, the kinetic energy approximately equal to the rest mass energy at the top of the barrier.

 The equation (\ref{ucrmod1}) agrees with the observed values of the critical velocity of a soliton moving over a barrier with an error of$\sim$ 3 percent.

Alternatively, we found that the energy is scaled up by a factor of $\sqrt{\tilde\lambda}$ at the top of the barrier, 

\begin{equation}
 \,M_{\,B}=\,M_{rest}\sqrt{\tilde\lambda}=\frac{8\sqrt{\tilde\lambda}}{3\sqrt{2}},
\end{equation}

 This agrees almost with the observed values. Now, the energy conservation in the non-relativistic limit is

 \begin{equation}
 \left( \frac{1}{2}\,u_{c}^{2}+1\right)\frac{8}{3\sqrt{2}} \approx \frac{8\sqrt{\tilde\lambda}}{3\sqrt{2}}
 \Rightarrow \,u_{c}^{2}\approx2\left(\sqrt{\tilde\lambda}-1\right),
 \end{equation}

where $\,u_{c}$ is the critical velocity. Hence, the critical velocity in the non-relativistic limit is 

\begin{equation}
 \,u_{c}\approx\sqrt{2\left(\sqrt{\tilde\lambda}-1\right)}
\end{equation}

Now, we will calculate the critical velocities for different barrier heights,{\it i.e}$\lambda_{0}>0$, using the above non-relativistic limit of the critical velocity,  and compare them the observed ones. Table 3 summarizes our calculations and observations of the critical velocities for different barrier heights.

\begin{table}
\begin{center}
\begin{tabular}{ccc} \\[6pt]
\hline
$\lambda_{0}$ & $\,u_{c}\left(non-relativistic\right)$ & observed $\,u_{c}$   \\ [6pt]
\hline
 0.125 &  0.34 &$ \sim 0.34$\\ 
 0.25 & 0.49 &  $\sim 0.45$\\
 0.5 & 0.67 &  $\sim 0.59 $ \\
 0.75  &  0.80 &  $\sim 0.65$  \\  [6pt]
\hline
\end{tabular}
\end{center}
\caption{The critical velocities in the non-relativistic limit vs the observed ones}
\end{table}

From table 3, one can see that for low barrier height and obviously low critical velocities there is an agreement between the calculated critical velocities and the numerically observed ones. However, for higher barrier heights there is a disagreement and the differences grow wider as we increase the height of the barrier. We can have a full agreement when we use the relativistic correction to the critical velocity.

 \begin{center}
  $\frac{\,M_{rest}}{\sqrt{1-u_{c}^{2}}}=\,M_{rest}\sqrt{\tilde\lambda}$\\
 $\Rightarrow \sqrt{1-u_{c}^{2}}=\frac{1}{\sqrt{\tilde\lambda}}.$
 \end{center}
Therefore, in the relativistic limit, the critical velocity reads

\begin{equation}
 \,u_{c}=\sqrt{\frac{\lambda_{0}}{1+\lambda_{0}}}.
\end{equation}

If we recalculate the critical velocities for the same barrier heights as in Tabel 3 we find an excellent agreement with numerically observed values, see table 4.

\begin{table}
\begin{center}
\begin{tabular}{ccc} \\[3pt]
\hline
$\lambda_{0}$ & $\,u_{c}\left(relativistic\right)$ & observed $\,u_{c}$   \\ [3pt]
\hline
 0.125 &  0.33 &$ \sim 0.34$\\ 
 0.25 & 0.45 &  $\sim 0.45$\\
 0.5 & 0.58&  $\sim 0.59 $ \\
 0.75  &  0.65 &  $\sim 0.65$  \\ [3pt]
\hline
\end{tabular}
\end{center}
\caption{The critical velocities in the relativistic limit vs the observed ones}
\end{table}

We were, also, able to produce the critical velocity curve using our approximation for the model by solving (\ref{eommod1}) using the forth order Runge-kutta method. Figure 7 shows the analytical critical velocity curve for a barrier of 0.25 height. The velocity of soliton which produces this curve is $\,u_{cr}=0.421025$ which is marginally less than the numerical value, {\it ie} $\,u_{cr}\sim0.45$. The system has an infinite degree of freedom and at the critical velocity the time of the interaction is large therefore this would contribute to the small difference between the two critical velocities. However, this is a further demonstration of the validity of our approximation.
\begin{figure}
\begin{center}
\includegraphics[angle=270, width=8cm]{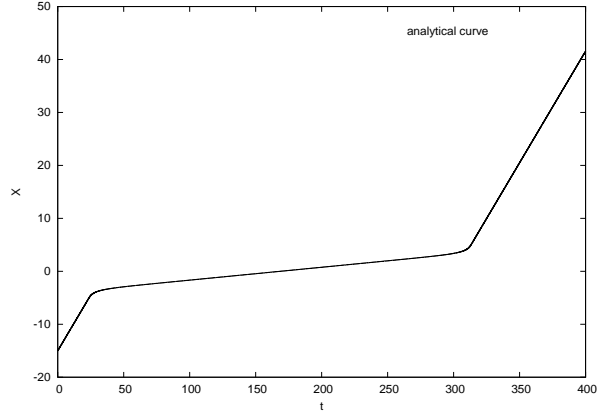}
\caption{The trajectory of soliton-barrier system showing the critical velocity in case of a barrier of 0.25 height}
\end{center}
\end{figure}

\section{Deformed Sine-Gordon Model}

We will analyse the scattering properties of  topological solitons in a class of model which is the generalisation of Sine-Gordon model and which has been recently proposed by Bazeia etl \cite{Bazeia}. The model depends on a positive real non-zero parameter $n$ but we will consider the model for its integer values as when $n=2$ the model reduces to Sine-Gordon one.

The model is constructed by a class of potentials which is given by

\begin{equation}
 \,\tilde V_{\,n}\left(\varphi\right)=\frac{2\tilde\lambda^{2}}{\,n^{2}}\,tan^{2}\left(\varphi\right)\left(1-\,sin^{\,n}\left(\varphi\right)\right),
\end{equation}

where the model will be approximated by the following soliton solutions
\begin{center}
 $\varphi=\,sin^{-1}\left[\,w\right]^{\frac{1}{n}}$\\   $\,w=\frac{\,exp\left[2\left(\,x-\,X\left(t\right)\right)\right]}{1+\,exp\left[2\left(\,x-\,X\left(t\right)\right)\right]}$.
\end{center}

The parameter $\tilde\lambda$ ,as before, is expressed in terms of the Haveiside function 


 The Lagrangian density is 

\begin{equation}
 \mathcal{L}_{\,n}=\frac{1}{2}\dot\varphi^{2}-\frac{1}{2}\grave{\varphi}^{2}-\,\tilde V_{\,n}\left(\varphi\right).
\end{equation}

Substituting the values of $\dot\phi, \phi^{\prime}$ and $\,\tilde V$, the Lagrangian density becomes

\begin{equation}
 \mathcal{L}_{\,n}=\frac{2}{\,n^{2}}\frac{\,w^{\frac{2}{\,n}}\left(1-\,w\right)^{2}}{\left(1-\,w^{\frac{2}{\,n}}\right)}\left(\,\dot X^{2}-1\right)-\frac{2\tilde\lambda^{2}}{\,n^{2}}\frac{\,w^{\frac{2}{\,n}}\left(1-\,w\right)^{2}}{\left(1-\,w^{\frac{2}{\,n}}\right)}.
\end{equation}

The Lagrangian is given by

\begin{eqnarray}\label{Lmod2}
 &&\,L_{\,n}=\int^{\infty}_{-\infty}\mathcal{L}_{\,n}\,d\,x.\nonumber\\
&&\quad\quad = \frac{2}{\,n^{2}}\left(\,\dot X^{2}-2\right)\int^{\infty}_{-\infty}\frac{\,w^{\frac{2}{\,n}}\left(1-\,w\right)^{2}}{\left(1-\,w^{\frac{2}{\,n}}\right)}\,d\,x\nonumber\\
&&-\frac{2\lambda_{0}}{\,n^{2}}\left(\lambda_{0}+2\right)\int^{\,x_{0}}_{-\,x_{0}}\frac{\,w^{\frac{2}{\,n}}\left(1-\,w\right)^{2}}{\left(1-\,w^{\frac{2}{\,n}}\right)}\,d\,x.
\end{eqnarray}

It is convenient to change variables
from $x$ to $r=\,w^{\frac{1}{n}}$. By applying this change of variables the first integral in the Lagrangian ({\it ie} (\ref{Lmod2})) becomes

\begin{equation}
 \int^{\infty}_{-\infty}\frac{\,w^{\frac{2}{\,n}}\left(1-\,w\right)^{2}}{\left(1-\,w^{\frac{2}{\,n}}\right)}=\frac{\,n}{2}\int^{1}_{0}\frac{\,r\left(1-\,r^{\,n}\right)}{\left(1-\,r^{2}\right)}\,d\,r.
\end{equation}

The Lagrangian can be recasted as

\begin{equation}
 \,L_{n}=\frac{1}{\,n}\left(\,\dot X^{2}-2\right)\,S_{n} -\lambda_{0}\left(\lambda_{0}+2\right)\mathcal{K}_{n},
\end{equation}

where  
\begin{equation}
\,S_{\,n}=\int^{1}_{0}\frac{\,r\left(1-\,r^{\,n}\right)}{1-\,r^{2}}\,dr, 
\end{equation}

\begin{equation}
\mathcal{K}_{n}=\frac{1}{n}\int^{\,x_{0}}_{-\,x_{0}}\frac{\,r\left(1-\,r^{\,n}\right)}{1-\,r^{2}}\,dr.
\end{equation}

After manipulating the integrals, We find that

\begin{equation}
\,S_{n}=\cases{ 1-ln2 & \,n=1 \cr
                               \frac{1}{2} & \,n=2. \cr}
\end{equation}

and all other satisfy the recurrence relation

\begin{equation}
 \,S_{n+2}=\,S_{n}+\frac{1}{n+2}.
\end{equation}

$\mathcal{K}_{n}$ will be expressed in terms of $\,w_{\pm}=\,w_{\pm}\left(\,X\right)$.

\begin{equation}
 \,w_{-}=\frac{\,exp\left(-2\left(\,X-\,x_{0}\right)\right)}{1+\,exp\left(-2\left(\,X-\,x_{0}\right)\right)},
\end{equation}

\begin{equation}
 \,w_{+}=\frac{\,exp\left(-2\left(\,X+\,x_{0}\right)\right)}{1+\,exp\left(-2\left(\,X+\,x_{0}\right)\right)}.
\end{equation}

 $\,n=1,..,6.$ are 

\begin{equation}
\mathcal{K}_{n}=\cases{ \left(\,w_{-}-\ln\left(1+\,w_{-}\right)-\left( \,w_{+}-\ln\left(1+\,w_{+}\right)\right)\right)&\,n=1 \cr
                                          \frac{1}{4}\left(\,w_{-}-\,w_{+}\right)& \,n=2 \cr
                                           \frac{1}{3}\left[\frac{\,w_{-}}{3}+\,w_{-}^{\frac{1}{3}}-\ln\left(1+\,w_{-}^{\frac{1}{3}}\right)-\left(\frac{\,w_{+}}{3}+\,w_{+}^\frac{1}{3}-\ln\left(1+\,w_{+}^{\frac{1}{3}}\right)\right)\right]&\,n=3\cr
                                         \frac{1}{4}\left[\frac{\,w_{-}^{\frac{1}{2}}}{2}+\frac{\,w_{-}}{4}-\left(\frac{\,w_{+}^{\frac{1}{2}}}{2}+\frac{\,w_{+}}{4}\right)\right]&\,n=4\cr
                                       \frac{1}{5}\left[\frac{\,w_{-}}{5}+\frac{\,w_{-}^{\frac{3}{5}}}{3}+\,w_{-}^{\frac{1}{5}}-\ln\left(1+\,w_{-}^{\frac{1}{5}}\right)-\left(\frac{\,w_{+}}{5}+\frac{\,w_{+}^{\frac{3}{5}}}{3}+\,w_{+}^{\frac{1}{5}}-\ln\left(1+\,w_{+}^{\frac{1}{5}}\right)\right)
                                         \right]&\,n=5\cr
                                       \frac{1}{6}\left[\frac{\,w_{-}^{\frac{2}{6}}}{2}+\frac{\,w_{-}^{\frac{4}{6}}}{4}+\frac{\,w_{-}}{6}-\left(\frac{\,w_{+}^{\frac{2}{6}}}{2}+\frac{\,w_{+}^{\frac{4}{6}}}{4}+\frac{\,w_{+}}{6}\right)\right]&\,n=6 \cr}
\end{equation}

The potential of the system is 
\begin{equation}
\,V_{n}\left(\,X\right)=\lambda_{0}\left(\lambda_{0}+2\right)\mathcal{K}_{n}
\end{equation}

The force on soliton by an obstruction is

\begin{equation}
 \,F_{n}\left(\,X\right)=-\lambda_{0}\left(\lambda_{0}+2\right)\frac{\partial\mathcal{K}_{n}}{\partial\,X}
\end{equation}

where

\begin{equation}
 -\frac{\partial\mathcal{K}_{n}}{\partial\,X}=\cases{ 2\left[\,w_{-}^{2}\left(\frac{1-\,w_{-}}{1+\,w_{-}}\right)-\left(\,w_{+}^{2}\left(\frac{1-\,w_{+}}{1+\,w_{+}}\right)\right)\right] & \,n=1 \cr
                                   \frac{1}{2}\left(\,w_{-}\left(1-\,w_{-}\right)-\,w_{+}\left(1-\,w_{+}\right)\right)& \,n=2 \cr
                                 \frac{2}{9}\,w_{-}^{\frac{1}{3}}\left(1-\,w_{-}\right)\left(1+\,w_{-}^{\frac{2}{3}}-\frac{1}{1+\,w_{-}^{\frac{1}{3}}}\right)&\cr
-\frac{2}{9}\,w_{+}^{\frac{1}{3}}\left(1-\,w_{+}\right)\left(1+\,w_{+}^{\frac{2}{3}}-\frac{1}{1+\,w_{+}^{\frac{1}{3}}}\right)&\,n=3\cr
                                     \frac{1}{8}\left[\,w_{-}\left(1-\,w_{-}\right)\left(1+\frac{1}{\,w_{-}^{\frac{1}{2}}}\right)-\left(\,w_{+}\left(1-\,w_{+}\right)\left(1+\frac{1}{\,w_{+}^{\frac{1}{2}}}\right)\right)\right]&\,n=4\cr
\frac{2}{25}\,w_{-}^{\frac{1}{5}}\left(1-\,w_{-}\right)\left(1+\,w_{-}^{\frac{4}{5}}+\,w_{-}^{\frac{2}{5}}-\frac{1}{1+\,w_{-}^{\frac{1}{5}}}\right)& \cr
\qquad-\frac{2}{25}\left(\,w_{+}^{\frac{1}{5}}\left(1-\,w_{+}\right)\left(1+\,w_{+}^{\frac{4}{5}}+\,w_{+}^{\frac{2}{5}}-\frac{1}{1+\,w_{+}^{\frac{1}{5}}}\right)\right)&\,n=5\cr
\frac{1}{18}\left[\,w{-}^{\frac{2}{6}}\left(1-\,w_{-}\right)\left(1+\,w_{-}^{\frac{2}{6}}+\,w_{-}^{\frac{4}{6}}\right)-\left(\,w_{+}^{\frac{2}{6}}\left(1-\,w_{+}\right)\left(1+\,w_{+}^{\frac{2}{6}}+\,w_{+}^{\frac{4}{6}}\right)\right)\right]&\,n=6 \cr}
\end{equation}
                               
 The Figures 10, 11 and the figures 12, 13 show the potentials and forces that Solitons in this model would experience at a barrier $\lambda_{0}=\pm1$ respectively.  
 The potentials and forces are decreasing as $\,n$ increasing. The potentials are getting more deformed, {\it ie} asymmetrical, as $\,n$ increases with the exception for the case $n=2$,({\it ie} the Sine-Gordon solution). 
\begin{figure}
\begin{center}
\includegraphics[angle=270, width=10cm]{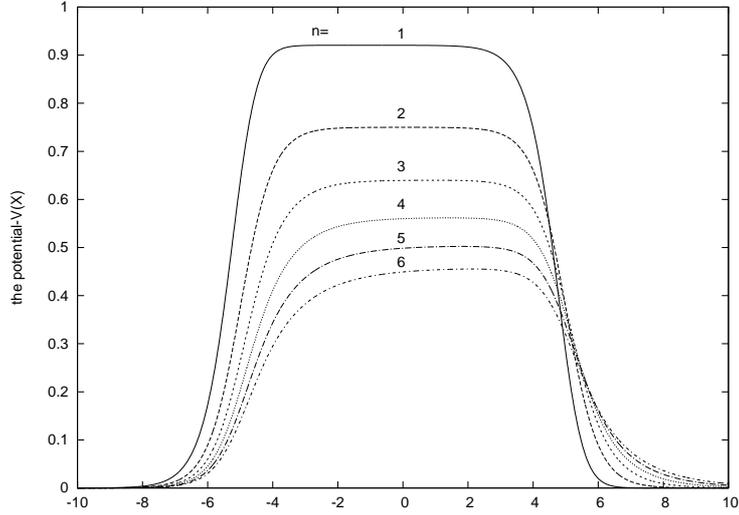}
\caption{The potentials raised by a barrier of height $\lambda_{0}=1$}
\end{center}
\end{figure}

\begin{figure}
\begin{center}
\includegraphics[angle=270, width=10cm]{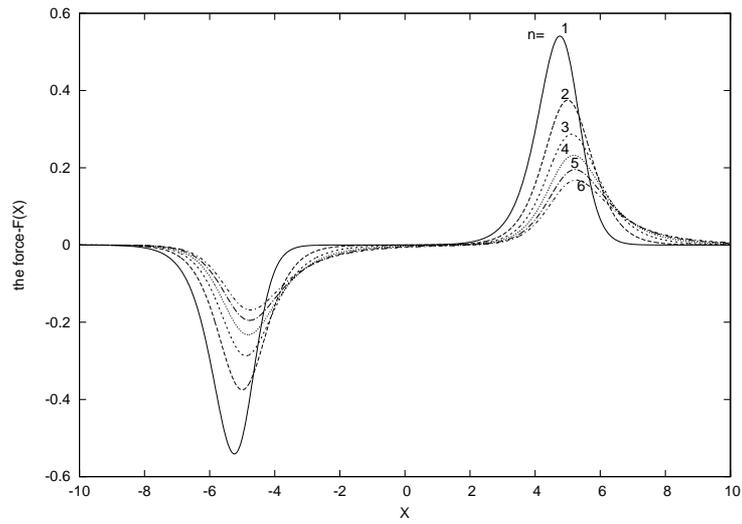}
\caption{The forces on solitons at a barrier of height $\lambda_{0}=1$}
\end{center}
\end{figure}

\begin{figure}
\begin{center}
\includegraphics[angle=270, width=10cm]{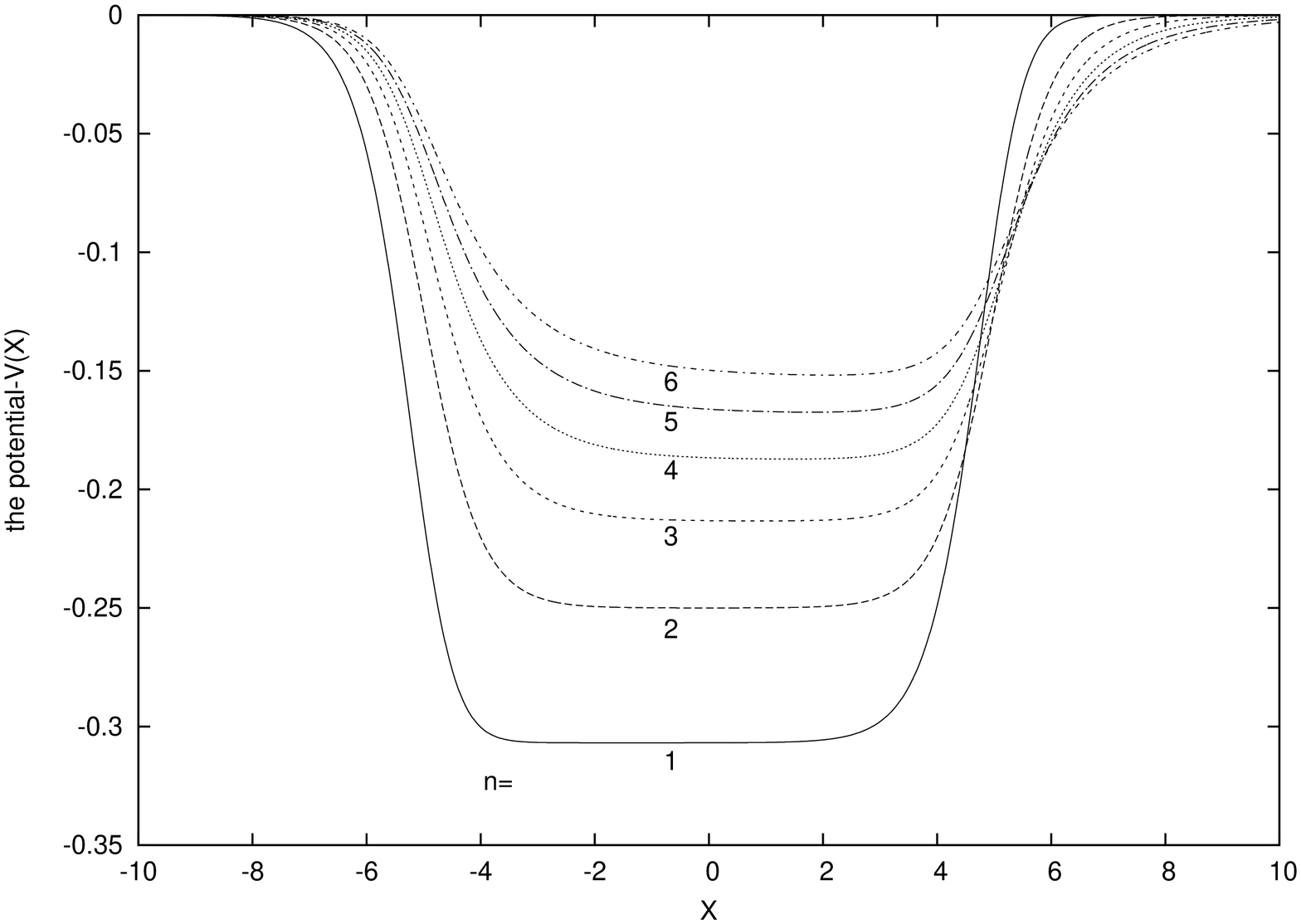}
\caption{The potentials raised by a hole,  $\lambda_{0}=-1$}
\end{center}
\end{figure}

\begin{figure}
\begin{center}
\includegraphics[angle=270, width=10cm]{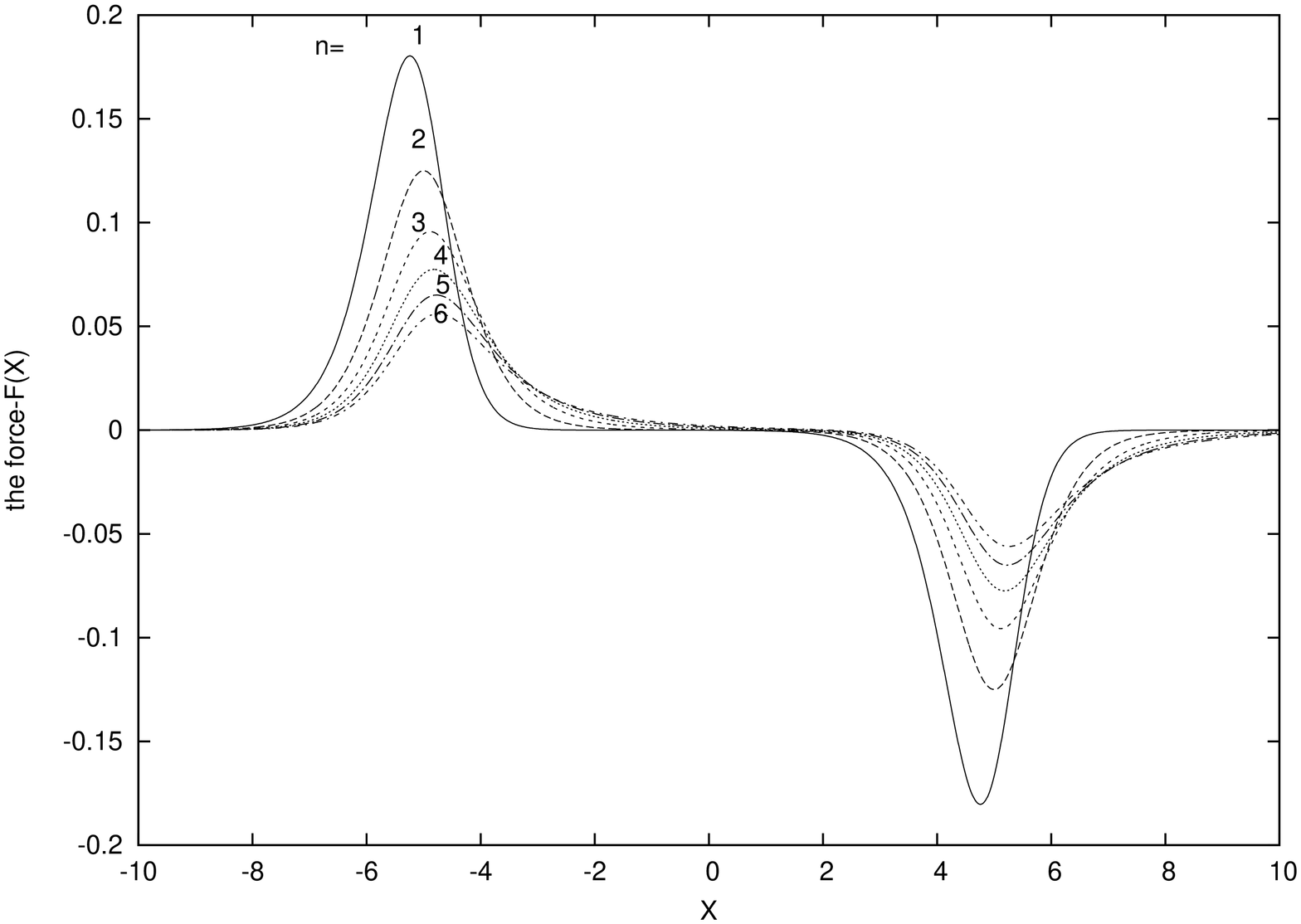}
\caption{The forces on solitons at a hole, $\lambda_{0}=-1$}
\end{center}
\end{figure}

Using the Euler-Lagrange equation, we can determine the equation of motion for each case. However, we will only consider the case where $\,n=2$ which is the Sine-Gordon solution.


The e.o.m for $\,n=2$ is 
\begin{equation}\label{eommod2}
\,\ddot X-\lambda_{0}\left(\lambda_{0}+2\right)\left(\,w_{-}\left(1-\,w_{-}\right)-\,w_{+}\left(1-\,w_{+}\right)\right)=0,
\end{equation}

and the rest mass for the Sine-Gordon soliton is $\, M_{rest(2)}=\frac{1}{2}$

The e.o.m. for $\,n=3$ is
\begin{eqnarray}
 &&\frac{2}{3}\left(\frac{4}{3}-\ln2\right)\,\ddot X - \frac{2}{9}\lambda_{0}\left(\lambda_{0}+2\right)\,w_{-}^{\frac{1}{3}}\left(1-\,w_{-}\right)\left(1+\,w_{-}^{\frac{2}{3}}-\frac{1}{1+\,w_{-}^{\frac{1}{3}}}\right)\nonumber\\
&&\qquad-\frac{2}{9}\lambda_{0}\left(\lambda_{0}+2\right)\left(\,w_{+}^{\frac{1}{3}}\left(1-\,w_{+}\right)\left(1+\,w_{+}^{\frac{2}{3}}-\frac{1}{1+\,w_{+}^{\frac{1}{3}}}\right)\right)=0.
\end{eqnarray}
 
and the rest mass for this soliton $\,M_{rest(3)}=\frac{2}{3}\left(\frac{4}{3}-\ln2\right)$.

When the soliton is far away from the obstruction, the equation of motion reduces to

\begin{equation}
 \,\ddot X=0, \qquad \,X=\,X_{0}+\,u\,t,
\end{equation}

where $\,X_{0}`$ is the initial position of the soliton.
The total energy 

\begin{eqnarray}
 &&\,E_{n}=\frac{\partial\,L_{n}}{\partial\,\dot X}\,\dot X-\,L_{n}\nonumber\\
&&=\frac{1}{n}\left(\,\dot X^{2}+2\right)\,S_{n}+\lambda_{0}\left(\lambda_{0}+2\right)\mathcal{K}_{n}
\end{eqnarray}

For $\,n=2$, the total energy is 

\begin{equation}
 \,E_{2}= \frac{1}{4}\left(\,\dot X^{2}+2\right)+\frac{\lambda_{0}}{4}\left(\lambda_{0}+2\right)\left(\,w_{-}-\,w_{+}\right)
\end{equation}

Far away from the obstruction, the energy is simply the 

\begin{equation}
 \frac{1}{4}\,\dot X^{2}+\frac{1}{2}.
\end{equation}

If the soliton is not moving, {\it ie} $\,\dot X=0$ then the total energy corresponds to the rest mass energy of the soliton. Therefore, 

\begin{equation}
 \,E_{2}\geq\frac{1}{2}
\end{equation}

The rest mass energy of the soliton at the top of the barrier({\it ie} $\lambda_{0}>0$) or inside a hole({\it ie}$ \lambda_{0}<0$) can be calculated by fixing the soliton position over the barrier or inside the hole. The best choice would be when the soliton is at $\,X=0$ because at this position the force on the soliton by the obstruction as can be seen from the figures is the minimum. In this case the functions $\,w_{\pm}=\,w_{\pm}\left(0\right)$ are 

\begin{equation}
 \,w_{-}\left(0\right)=\frac{\,exp\left(2\left(\,x_{0}\right)\right)}{1+\,exp\left(2\left(\,x_{0}\right)\right)},
\end{equation}
 
\begin{equation}
 \,w_{+}\left(0\right)=\frac{\,exp\left(-2\left(\,x_{0}\right)\right)}{1+\,exp\left(-2\left(\,x_{0}\right)\right)}.
\end{equation}

Since $\,x_{0}=5.$ this would give with a good approximation $\,w_{-}\approx1$ and $\,w_{+}\approx0$.

 Setting $\,\dot X=0$ at the obstruction the solitons masses are  given by
\begin{equation}
 \,M_{n}=\,M_{rest}+\frac{\,M_{rest}}{2}\lambda_{0}\left(\lambda_{0}+2\right)
\end{equation}

Hence, the masses of solitons at the obstruction for $\,n=1,..,6.$ are 

\begin{equation}
 \,M_{n}=\cases{ 0.614+0.307\lambda_{0}\left(\lambda_{0}+2\right)&\,n=1 \cr
                       0.500+0.25\lambda_{0}\left(\lambda_{0}+2\right)& \,n=2 \cr
                        0.427+0.2135\lambda_{0}\left(\lambda_{0}+2\right)& \,n=3 \cr
                        0.375+0.1875\lambda_{0}\left(\lambda_{0}+2\right)&\,n=4 \cr
                        0.336+0.168\lambda_{0}\left(\lambda_{0}+2\right)& \,n=5 \cr
                       0.306+0.153\lambda_{0}\left(\lambda_{0}+2\right)& \,n=6 \cr}
\end{equation}

Table 5 and table 6  compare between the numerical observed masses and the calculated ones for a barrier of 0.4 height and a hole of -0.4 deep.

\begin{table}
\begin{center}
\begin{tabular}{ccc} 
\multicolumn{3}{c}{$\lambda_{0}=0.4$}\\ [3pt]
\hline
$\,n$ & calculated $\,M_{B}$ & observed $\,M_{B}$ \\ [3pt]
\hline
 1 &0.908& 0.908\\
 2&  0.74& 0.74\\
 3  & 0.632 &0.631 \\
4 &0.555 & 0.554 \\
5 &  0.497& 0.496 \\
6 &0.453 & 0.449 \\ [3pt]
\hline
\end{tabular}
\end{center}
\caption{Solinton masses at the top of a barrier of 0.4 height}
\end{table}

\begin{table}
\begin{center}
\begin{tabular}{ccc} 
\multicolumn{3}{c}{$\lambda_{0}=-0.4$}\\ [3pt]
\hline
$\,n$ & calculated $\,M_{H}$ & observed $\,M_{H}$  \\ [3pt]
\hline
 1 &0.417 &0.417   \\
 2&0.34& 0.3400  \\
 3& 0.2904  & 0.2904 \\
4 & 0.255 & 0.255  \\
5 & 0.2016 & 0.2297 \\
6 &0.1836 & 0.2097\\  [3pt]
\hline
\end{tabular}
\end{center}
\caption{Soliton masses inside a hole of -0.4 deep}
\end{table}

Thus, the agreement between the numerical and calculated values for the masses at the top of a barrier or inside a hole is perfect. This shows the validity of our approximation.


The critical velocities for solitons-barrier system can be calculated easily as before using the following equation

\begin{eqnarray}\label{ucrmod2}
 &&\,u_{c}=\sqrt{1-\left(\frac{\,M_{rest}}{\,M_{B}}\right)^{2}}\nonumber \\
&&\qquad = \sqrt{1-\left(\frac{2}{2+\lambda_{0}\left(\lambda_{0}+2\right)}\right)^{2}}
\end{eqnarray}
 For a barrier of height 0.4, the critical velocity obtained numerically for $,n=1,..6$ solutions is $\sim 0.7$ \cite{AZ08}. Using the above equation, the critical velocity is 0.737. The $\sim 5.3$ percent difference is because the dynamics of this system  is very sensitive to the perturbation raised by the obstruction. Unlike the $\lambda\varphi^{4}$ model, in this class of models solitons are excited as they meet the obstruction in their way and this would contribute to the masses of solitons as we have discussed that in \cite{AZ08}. If we subtract the excitation energy from the masses observed numerically as we did in \cite{AZ08} then there will be an excellent agreement between them.

We will solve the equation of motion for $\,n=2$ soliton solution ({\it ie} (\ref{eommod2}) using forth order Runge-Kutta method with initial conditions that specify the speed of the soliton and its position. 


 Fig. 14 shows a good agreement between the analytical and numerical simulation curves for $\,n=2$ soliton moving with a speed of 0.45 and encountering a hole of -0.1 depth. However, fig. 15 shows that the there is a noticeable difference between the analytical and the numerical curves for $\,n=2$ soliton moving with a speed of 0.5 and meeting a barrier of height 0.1. We found out that as we are getting closer to the critical velocity of the system the difference between the analytical and the numerical simulation curves grow wider. In the case of a barrier of 0.1 the critical velocity is 0.425 and  if we keep increasing the velocity the difference is diminishing. The curves in the case of the hole, fig. 14, the speed of soliton is much larger than the critical velocity and so the curves are in a good agreement. 

\begin{figure}
\begin{center}
\includegraphics[angle=270, width=8cm]{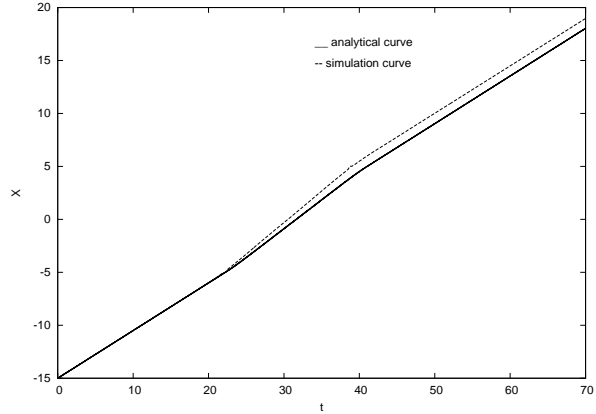}
\caption{The trajectories for $\,n=2$ soliton solution over a barrier of 0.1 height, $\,u=0.45$}
\end{center}
\end{figure}

\begin{figure}
\begin{center}
\includegraphics[angle=270, width=8cm]{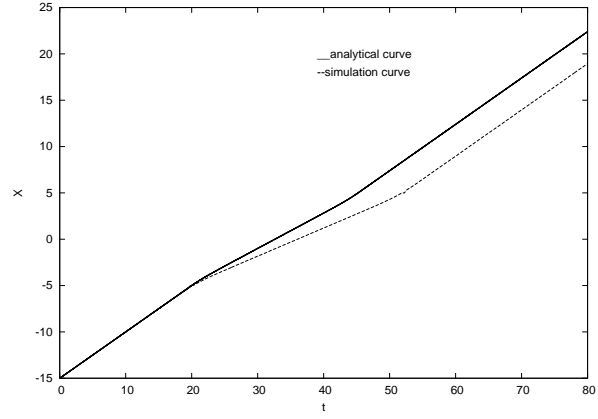}
\caption{The trajectories for $\,n=2$ soliton solution over a barrier of 0.1 height, $\,u=0.45$, $\alpha=1.40$ }
\end{center}
\end{figure}

For a barrier of 0.4 height, the critical velocity for $\,n=2$ solution as calculated from (\ref{ucrmod2}) is $\sim0.7$ which agrees with numerical value. By using the fourth order Lunge-kutta method in solving the equation of motion for $\,n=2$ we found out that the critical curve can be obtained with a critical velocity, $\,u_{c}=0.692375$ which is in perfect agreement with the numerical value. Fig. 16 shows the  critical trajectory in case of a barrier of 0.4 height.

 \begin{figure}
\begin{center}
\includegraphics[angle=270, width=8cm]{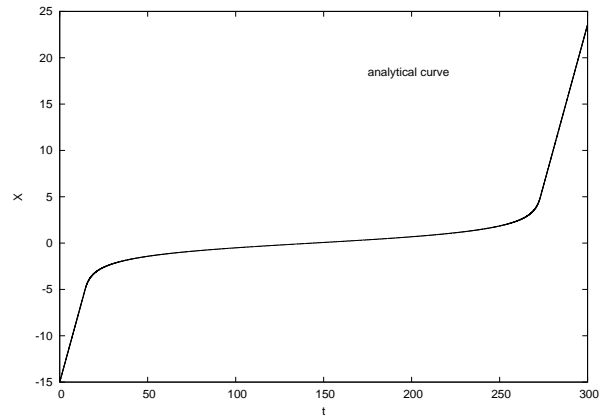}
\caption{The critical trajectory for $\,n=2$ soliton in case of a barrier of 0.4 height, $\,u_{c}=0.692375$  }
\end{center}
\end{figure}

\section{ Q-ball-Obstruction System}

Consider the field configuration

\begin{equation}
 \Phi=\eta\left(\,t\right)\,e^{\,i\theta\left(\,t\right)}\,f\left(\,x-\,X\left(\,t\right)\right),
\end{equation}

where $\eta$ and $\theta$ are time-dependent real moduli and $\,f\left(\,x-\,X\left(\,t\right)\right)$ is the profile function, which does not have any explicit dependence on $\,t$, of a Q-ball with frequency $\omega$. We will study the dynamics of the system without any obstruction, {\it ie} $\tilde\lambda=1$. The Lagrangian density in $\left(1+1\right)$ dimensions of our model is 

\begin{equation}
 \mathcal{\,L}=\frac{1}{2}\vert\dot{\Phi}\vert^{2}-\frac{1}{2}\vert\Phi_{\,x}\vert^{2}-2\vert\Phi\vert^{2}+2\vert\Phi\vert^{4}-\vert\Phi\vert^{6},
\end{equation}

Each term in this Lagrangian density is given in the following equations

\begin{equation}
 \vert\Phi\vert^{2}=\eta^{2}\,f^{2},
\end{equation}

\begin{equation}
 \vert\Phi\vert^{4}=\eta^{4}\,f^{4},
\end{equation}

\begin{equation}
 \vert\Phi\vert^{6}=\eta^{6}\,f^{6},
\end{equation}

\begin{eqnarray}
&& \dot{\Phi}=\dot{\eta}\,e^{\,i\theta}\,f+\,i\dot{\theta} \,e^{\,i\theta}\,f+\eta\,e^{\,i\theta}\,f_{\,X}\dot{\,X},\nonumber\\
&&\Rightarrow\vert\dot{\Phi}\vert^{2}=\left(\dot{\eta}^{2}+\eta^{2}\dot{\theta}^{2}\right)\,f^{2}+\eta^{2}\dot{\,X}^{2}\,f^{2}_{\,X},
\end{eqnarray}

\begin{eqnarray}
&&\Phi_{\,x}=\eta\,e^{\,i\theta}\,f_{\,x},\nonumber\\
&&\Rightarrow\vert\Phi_{\,x}\vert^{2}=\eta^{2}\,f_{\,x}^{2}.
\end{eqnarray}

The Lagrangian is 

\begin{equation}
 \,L=\int\,d\,x\mathcal{\,L}.
\end{equation}

The Lagrangian after inserting each term of the Lagrangian density is 

\begin{equation}
 \,L=\frac{1}{2}\left(\dot{\eta}^{2}+\eta^{2}\dot{\theta}^{2}\right)\,I_{2}+\frac{1}{2}\eta^{2}\dot{\,X}^{2}\,f^{2}_{\,X}-\frac{1}{2}\eta^{2}\,I_{\,x}-2\eta^{2}\,I_{2}+2\eta^{4}\,I_{4}-\eta^{6}\,I_{6},
\end{equation}

where

\begin{equation}
 \,I_{\,n}=\int^{\infty}_{-\infty}\,f^{\,n}\,d\,x, \quad \,n=2,4,6
\end{equation}

\begin{equation}
 \,I_{\,x}=\int^{\infty}_{-\infty}\,f_{\,x}^{2}\,d\,x,
\end{equation}

\begin{equation}
 \,I_{\,X}=\int^{\infty}_{-\infty}\,f_{\,X}^{2}\,d\,x.
\end{equation}

Applying the Euler-Lagrange equations 

\begin{equation}
\frac{\partial\,L}{\partial\eta}-\frac{\,d}{\,d\,t}\left(\frac{\partial\,L}{\partial\dot{\eta}}\right)=0,
\end{equation}
\begin{equation}
 \frac{\partial\,L}{\partial\theta}-\frac{\,d}{\,d\,t}\left(\frac{\partial\,L}{\partial\dot{\theta}}\right)=0,
\end{equation}

\begin{equation}
 \frac{\partial\,L}{\partial\,X}-\frac{\,d}{\,d\,t}\left(\frac{\partial\,L}{\partial\dot{\,X}}\right)=0,
\end{equation}

 yields the following equations of motion 

\begin{equation}
 \left(\ddot{\eta}-\eta\dot{\theta}^{2}+4\eta\right)\,I_{2}-\eta\dot{\,X}^{2}\,I_{\,X}+\eta\,I_{\,x}-8\eta^{3}\,I_{4}+6\eta^{5}\,I_{6}=0
\end{equation}

\begin{equation}
 \ddot{\theta}+2\left(\frac{\dot{\eta}}{\eta}\right)\dot{\theta}=0
\end{equation}

\begin{equation}
\ddot{\,X}+2\left(\frac{\dot{\eta}}{\eta}\right)\dot{\,X}=0.
\end{equation}

We can have a conserved quantity using

\begin{equation}
 \frac{\,d}{\,d\,t}\left(\frac{\partial\,L}{\partial\dot\theta}\right) \Rightarrow \frac{\partial\,L}{\partial\dot\theta}=const.
\end{equation}
from which we obtain

\begin{equation}
\eta^{2}\dot\theta=const.
\end{equation}

Similarly, 

\begin{equation} 
\frac{\,d}{\,d\,t}\left(\frac{\partial\,L}{\partial\dot{\,X}}\right)\Rightarrow\frac{\partial\,L}{\partial\dot{\,X}}=const.
\end{equation}
from which we obtain

\begin{equation}
\eta^{2}\dot{\,X}=Const.
\end{equation}

If $\dot{\eta}=0$, that is when the magnitude of the Q-ball does not change with time we have

\begin{equation}
 \ddot{\,X}=0 \Rightarrow\,X-\,X_{0}=\,u\,t,
\end{equation}

where $\,u$ is the velocity of the Q-ball. Also,

\begin{equation}
 \ddot{\theta}=0\Rightarrow\theta-\theta_{0}=\omega\,t.
\end{equation}

The energy density

\begin{equation}
 \mathcal{\,H}=\frac{\partial\mathcal{\,L}}{\partial\dot{\Phi}}\dot{\Phi}+\frac{\partial\mathcal{\,L}}{\partial\dot{\Phi}^{\dagger}}\dot{\Phi}^{\dagger}-\mathcal{\,L}
\end{equation}




The total energy 

\begin{eqnarray}
&&\,E=\int^{\infty}_{-\infty}\mathcal{\,H}\,d\,x \nonumber\\
&&=\frac{1}{2}\left(\dot{\eta}^{2}+\eta^{2}\dot{\theta}^{2}\right)\,I_{2}+\frac{1}{2}\eta^{2}\dot{\,X}^{2}\,I_{\,X}+\frac{1}{2}\eta^{2}\,I_{\,x}\nonumber\\
&&+2\eta^{2}\,I_{2}-2\eta^{2}\,I_{2}-2\eta^{4}\,I_{4}+\eta^{6}\,I_{6}
\end{eqnarray}

If we assume that $\eta=1$ and there is no change of this amplitude as the time evolves, $\dot{\eta}=0$ then the total energy becomes

\begin{equation}
 \,E=\frac{1}{2}\dot{\theta}^{2}\,I_{2}+\frac{1}{2}\dot{\,X}^{2}\,I_{\,X}+\frac{1}{2}\,I_{\,x}+2\,I_{2}-2\,I_{2}-2\,I_{4}+\,I_{6}
\end{equation}

If the soliton is not moving ( {\it ie}$\dot{\,X}=0$) then the total energy reduces to the rest energy of the soliton

\begin{equation} \label{Erest}
 \,E_{rest}=\frac{1}{2}\dot\theta^{2}\,I_{2}+\frac{1}{2}\,I_{\,x}+2\,I_{2}-2\,I_{2}-2\,I_{4}+\,I_{6}
\end{equation}

Now, let us introduce an obstruction to the system ({\it ie} $\lambda_{0}\neq0$) and look over the dynamics of the system.  Let us assume that the scattering with an obstruction that leaves the Q-ball stable will not affect the magnitude of the Q-ball, {\it ie} $\eta\left(t\right)\approx 1$. Thus, we can have the field configuration approximated by the phase and position parameters. Thus , we have

\begin{equation}
 \Phi\left(\,x-\,X\left(t\right)\right)= \,e^{\,i\theta\left(\,t\right)}\,f\left(\,x-\,X\left(\,t\right)\right).
\end{equation}

The Lagrangian density with an obstruction introduced to the system is
\begin{equation}
\mathcal{\,L}=\frac{1}{2}\vert\dot{\Phi}\vert^{2}-\frac{1}{2}\vert\Phi_{\,x}\vert^{2}-\tilde\lambda\left(2\vert\Phi\vert^{2}-2\vert\Phi\vert^{4}+\vert\Phi\vert^{6}\right).
\end{equation}

Now, each term in this Lagrangian density is given by

\begin{equation}
 \vert\dot\Phi\vert^{2}=\dot\theta^{2}\,f^{2}+\,\dot X^{2}\,f^{2}_{\,X},
\end{equation}

\begin{equation}
 \vert\Phi_{\,x}\vert^{2}=\,f_{\,x}^{2},
\end{equation}

\begin{equation}
 \vert\Phi\vert^{\,n}=\,f^{\,n}, \quad \,n=2,4,6.
\end{equation}
The Lagrangian density becomes

\begin{equation}
 \mathcal{L}= \frac{1}{2}\dot\theta^{2}\,f^{2}+\frac{1}{2}\,\dot X^{2}\,f^{2}_{\,X}-\frac{1}{2}\,f_{\,x}^{2}-\tilde\lambda\left(2\,f^{2}-2\,f^{4}+\,f^{6}\right)
\end{equation}

The Lagrangian is 

\begin{eqnarray}
&& \,L=\int^{\infty}_{-\infty}\,d\,x\mathcal{L}\nonumber\\
&&=\frac{1}{2}\dot\theta^{2}\,I_{2}+\frac{1}{2}\dot{\,X}^{2}\,I_{\,X}-\frac{1}{2}\,I_{\,x}-2\,I_{2}+2\,I_{4}-\,I_{6}-\lambda_{0}\left(2\,I^{\prime}_{2}-2\,I^{\prime}_{4}+\,I^{\prime}_{6}\right),
\end{eqnarray}

where

\begin{equation}
 \,I^{\prime}_{n}=\int^{\,x_{0}}_{-\,x_{0}}\,d\,x\,f^{n}, \quad \,n=2,4,6.
\end{equation}

The total energy is 

\begin{eqnarray}
&& \,E=\frac{\partial\,L}{\partial\,\dot X}\,\dot X+\frac{\partial\,L}{\partial\dot\theta}\dot\theta-\,L\nonumber\\
&&=\frac{1}{2}\,\dot X^{2}\,I_{\,X}+\frac{1}{2}\dot\theta^{2}\,I_{2}+\frac{1}{2}\,I_{\,x}+2\,I_{2}-2\,I_{4}+\,I_{6}\nonumber\\
&&\qquad+\lambda_{0}\left(2\,I^{\prime}_{2}-2\,I^{\prime}_{4}+\,I^{\prime}_{6}\right).
\end{eqnarray}

In case the Q-ball is far away from obstruction, the total energy reduces to (\ref{Erest}).

The field equations after using Euler-Lagrange equations for the parameters $\,X$ and $\theta$ are

\begin{equation}\label{Qfield}
 \,\ddot X\,I_{\,X}+\lambda_{0}\left[2\frac{\partial,I^{\prime}_{2}}{\partial\,X}-2\frac{\partial,I^{\prime}_{4}}{\partial\,X}+\frac{\partial,I^{\prime}_{6}}{\partial\,X}\right]=0
\end{equation}

\begin{equation}
 \ddot\theta=0 \Rightarrow \dot\theta=constant.
\end{equation}

$\dot\theta$ corresponds to the angular frequency of the Q-ball, $\dot\theta=\omega$. 

 There is an exact solution to such system in (1+1) dimension \cite{AZ09}. Let us approach the dynamics of the system by using an approximate solution given by

\begin{equation}
 \,f\left(\,x-\,X\left(t\right)\right)=\left[\frac{4-\omega^{2}}{2+\sqrt{2\omega^{2}-4}\,\cosh\left(2\sqrt{4-\omega^{2}}\left(\,x-\,X\left(t\right)\right)\right)}\right]^{\frac{1}{2}},
\end{equation}

where $\omega=\dot\theta$. The potential of our model put restrictions on the values of $\omega$ for which a Q-ball solution exists. Thus , the Q-ball field exist for $\omega$ in this range, {\it ie} $\sqrt{2}<\omega<2$. We will select a particular phase that simplify our model and then look over the dynamics of the soliton-obstruction system for this phase. We will choose, for analytical simplicity, $\omega=\sqrt{3}$ and the approximate solution reduces to

\begin{equation}
 \,f\left(\,x-\,X\left(t\right)\right)=\sqrt{\frac{1}{2+\sqrt{2}\,cosh\left(2\left(\,x-\,X\left(t\right)\right)\right)}}.
\end{equation}

 We will evaluate the integrals,$,I_{\,X}, \,I_{\,x}$, and $\,I_{n}$ using this solution. Thus, we have

\begin{equation}
\,I_{2}=\int^{\infty}_{-\infty}\,f^{2}\,d\,x=0.623,
\end{equation}

\begin{equation}
 \,I_{4}=\int^{\infty}_{-\infty}\,f^{4}\,d\,x=0.123,
\end{equation}

\begin{equation}
 \,I_{6}=\int^{\infty}_{-\infty}\,f^{6}\,d\,x=0.029,
\end{equation}

\begin{equation}
\,I_{\,x}=\int_{-\infty}^{\infty}\,f_{\,x}^{2}\,dx=0.1885,
\end{equation}

\begin{equation}
\,I_{\,x}=\int_{-\infty}^{\infty}\,f_{\,X}^{2}\,dx=0.1885,
\end{equation}

Calculating the energy density away from the obstruction for a static Q-ball, (\ref{Qfield}), with $\omega=\sqrt{3}$ using the above values of the integrals gives $\,E=2.058$ which is the rest mass of the Q-ball and this is in agreement with the numerical value.

Now, the Lagrangian becomes
\begin{equation}
 \,L=\frac{1}{2}\left(0.1885\right)\left(\,\dot X^{2}-2\right)-\lambda_{0}\left(2\,I^{\prime}_{2}-2\,I^{\prime}_{4}+\,I^{\prime}_{6}\right),
\end{equation}

We will evaluate the integrals $\,I^{\prime}_{n}, n=2,4,6$.

For $n=2$

\begin{eqnarray}
 &&\,I^{\prime}_{2}=\frac{1}{\sqrt{2}}\,tanh^{-1}\left(0.4142\,tanh\left[\left(\,X+\,x_{0}\right)\right]\right)\nonumber\\
&&\qquad-\frac{1}{\sqrt{2}}\,tanh^{-1}\left(0.4142\,tanh\left[\left(\,X-\,x_{0}\right)\right]\right).
\end{eqnarray}

For $n=4$, 

\begin{eqnarray}
&& \,I^{\prime}_{4}=\nonumber\\
&&\frac{1}{\sqrt{2}}\,tanh^{-1}\left[0.4142\,tanh\left(\,X+\,x_{0}\right)\right]-\frac{\,sinh\left(2\left(\,X+\,x_{0}\right)\right)}{4\sqrt{2}+4\,cosh\left(2\left(\,X+\,x_{0}\right)\right)}\nonumber\\
&&-\frac{1}{\sqrt{2}}\,tanh^{-1}\left[0.4142\,tanh\left(\,X-\,x_{0}\right)\right]+\frac{\,sinh\left(2\left(\,X-\,x_{0}\right)\right)}{4\sqrt{2}+4\,cosh\left(2\left(\,X-\,x_{0}\right)\right)}
\end{eqnarray}

For $n=6$

\begin{eqnarray}
&& \,I^{\prime}_{6}=\nonumber\\
&&=\frac{5tanh^{-1}\left(0.4142tanh\left(X+x_{0}\right)\right)\left[8cosh\left(2\left(X+x_{0}\right)\right)+\sqrt{2}cosh\left(4\left(X+x_{0}\right)\right)+5\sqrt{2}\right]}{8\left[2+\sqrt{2}cosh\left(4\left(X+x_{0}\right)\right)\right]^{2}}\nonumber\\
&&-\frac{7\sqrt{2}sinh\left(2\left(X+x_{0}\right)\right)-3sinh\left(4\left(X+x_{0}\right)\right)}{8\left[2+\sqrt{2}cosh\left(2\left(X+x_{0}\right)\right)\right]^{2}}\nonumber\\
&&-\frac{5tanh^{-1}\left(0.4142tanh\left(X-x_{0}\right)\right)\left[8cosh\left(2\left(X-x_{0}\right)\right)+\sqrt{2}cosh\left(4\left(X-x_{0}\right)\right)+5\sqrt{2}\right]}{8\left[2+\sqrt{2}cosh\left(4\left(X-x_{0}\right)\right)\right]^{2}}\nonumber\\
&&+\frac{7\sqrt{2}sinh\left(2\left(X-x_{0}\right)\right)+3sinh\left(4\left(X-x_{0}\right)\right)}{8\left[2+\sqrt{2}cosh\left(2\left(X-x_{0}\right)\right)\right]^{2}}.
\end{eqnarray}

The potential is

\begin{equation}
 \,V\left(\,X\right)=\lambda_{0}\left(2\,I^{\prime}_{2}-2\,I^{\prime}_{4}+\,I^{\prime}_{6}\right).
\end{equation}

Fig. 15 and fig. 16 show the potential raised by a barrier of height, $\lambda_{0}=1$ and by a hole of $\lambda_{0}=-1$ depth located between $\vert\,x_{0}\vert\leq10$.

\begin{figure}
\begin{center}
\includegraphics[angle=270, width=10cm]{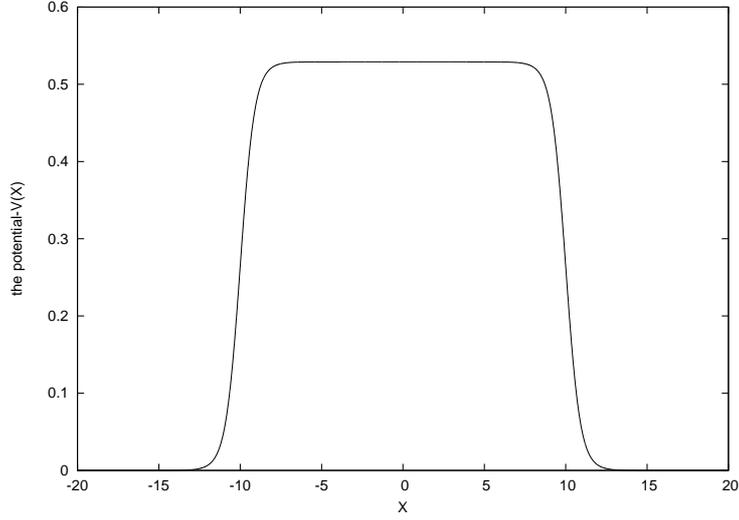}
\caption{The potentials raised by a barrier, $\lambda_{0}=1$}
\end{center}
\end{figure}

\begin{figure}
\begin{center}
\includegraphics[angle=270, width=10cm]{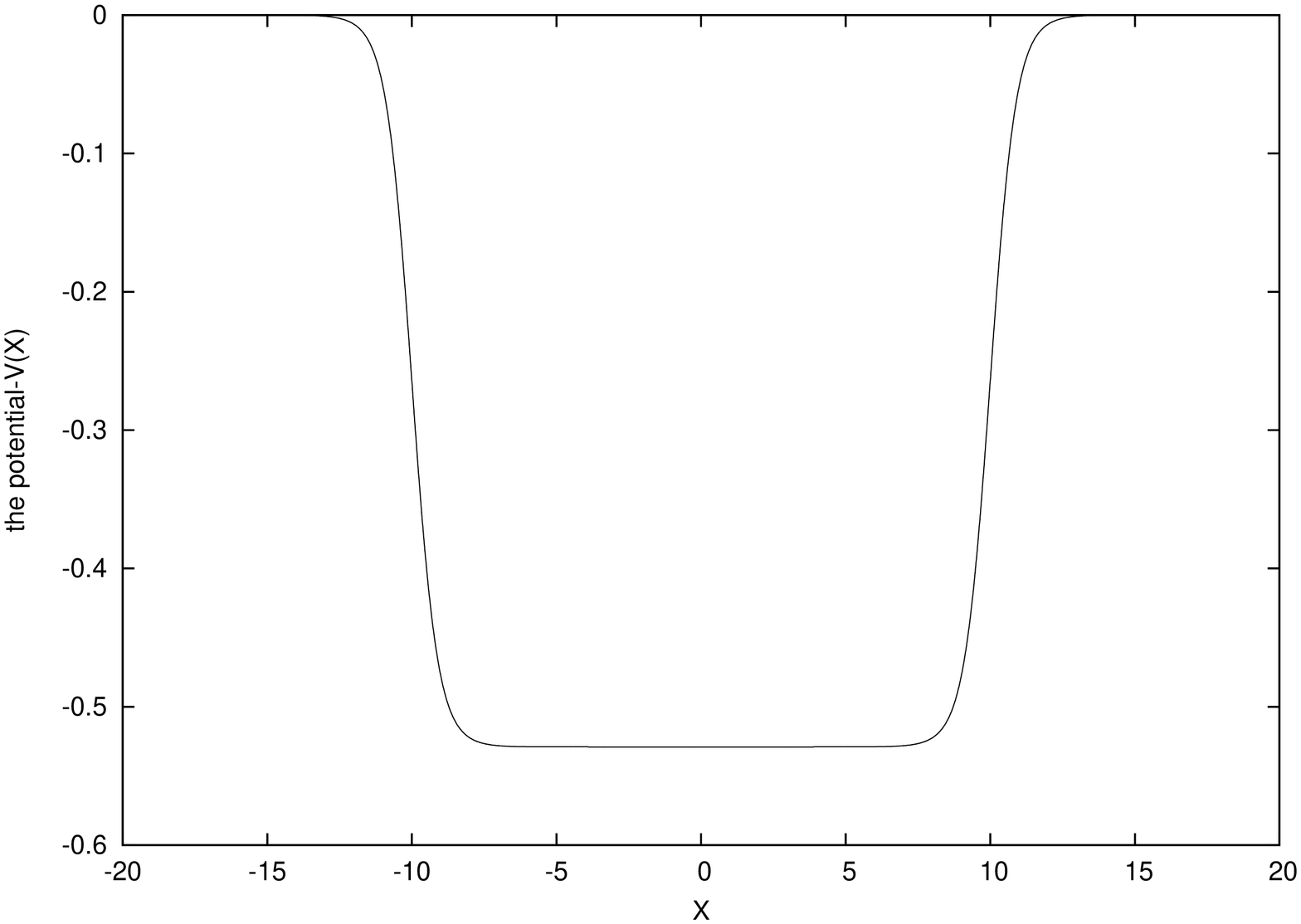}
\caption{The potential raised by a hole, $\lambda_{0}=-1$}
\end{center}
\end{figure}

The force between the Q-ball and the obstruction is 

\begin{eqnarray}
&& \,F\left(\,X\right)=-\frac{\partial\,V}{\partial\,X}\nonumber\\
&& = -\lambda_{0}\left[2\frac{\partial\,I^{\prime}_{2}}{\partial\,X}-2\frac{\partial\,I^{\prime}_{4}}{\partial\,X}+\frac{\partial\,I^{\prime}_{6}}{\partial\,X}\right].
\end{eqnarray}

where 

\begin{eqnarray}
&& \frac{\partial\,I^{\prime}_{2}}{\partial\,X}\nonumber\\
&&=\frac{1}{2+\sqrt{2}\,cosh\left(2\left(\,X+\,x_{0}\right)\right)}-\frac{1}{2+\sqrt{2}\,cosh\left(2\left(\,X-\,x_{0}\right)\right)},
\end{eqnarray}

\begin{eqnarray}
 && \frac{\partial\,I^{\prime}_{4}}{\partial\,X}\nonumber\\
&&=\frac{1}{\left(2+\sqrt{2}\,cosh\left(2\left(\,X+\,x_{0}\right)\right)\right)^{2}}-\frac{1}{\left(2+\sqrt{2}\,cosh\left(2\left(\,X-\,x_{0}\right)\right)\right)^{2}},
\end{eqnarray}

\begin{eqnarray}
 && \frac{\partial\,I^{\prime}_{6}}{\partial\,X}\nonumber\\
&&=\frac{1}{\left(2+\sqrt{2}\,cosh\left(2\left(\,X+\,x_{0}\right)\right)\right)^{3}}-\frac{1}{\left(2+\sqrt{2}\,cosh\left(2\left(\,X-\,x_{0}\right)\right)\right)^{3}}.
\end{eqnarray}

Fig. 17 and fig. 18  show the force on the Q-ball due to barrier and a hole.
\begin{figure}
\begin{center}
\includegraphics[angle=270, width=10cm]{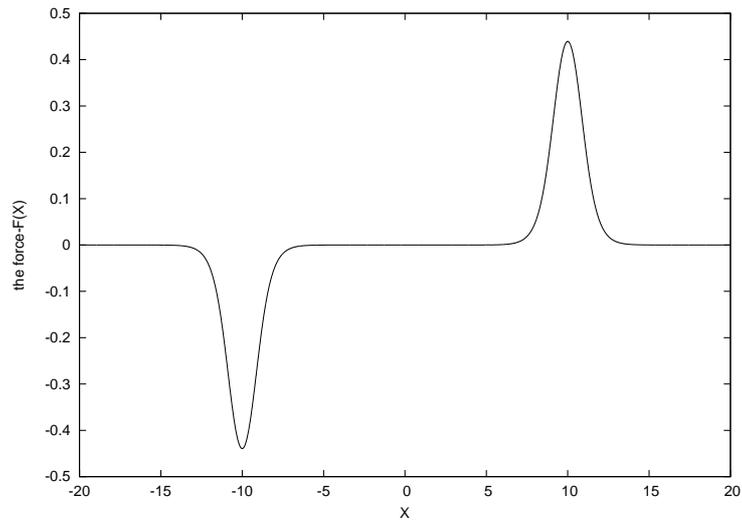}
\caption{The force on the Q-ball by a barrier, $\lambda_{0}=1$}
\end{center}
\end{figure}

\begin{figure}
\begin{center}
\includegraphics[angle=270, width=10cm]{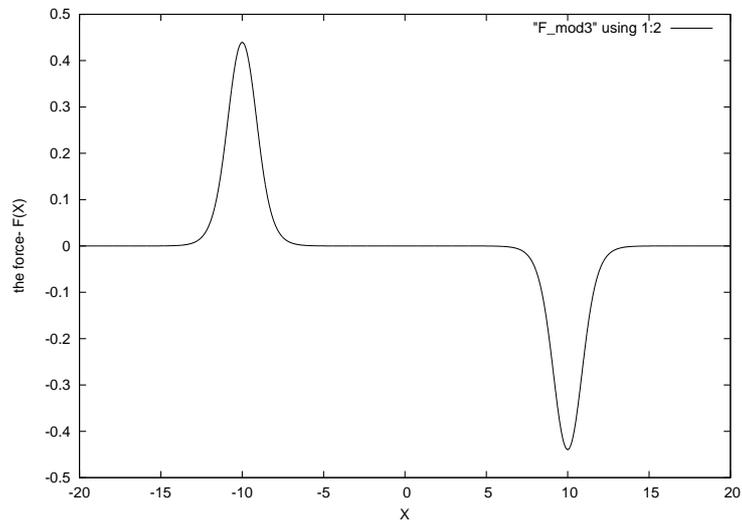}
\caption{The force on the Q-ball by a hole, $\lambda_{0}=-1$}
\end{center}
\end{figure}

As we have discussed in the previous chapter, the stability of the Q-ball is affected by the a deep potential hole while it is not affected for the case of high barriers. So, we considered only shallow holes. We will calculate the rest energy for a static Q-ball with $\omega=\sqrt{3}$, {\it ie} $\,\dot X=0$,  at the top of a barrier and inside a hole using (84) and compare it to the numerical values. To perform our calculation we fixed the position of soliton at the obstruction and a good choice as before is at $\,X=0$ where the interaction with obstruction is minimized. The rest energy at the obstruction is given by 

\begin{equation}\label{EQ}
 \,E=\,E_{rest}+\lambda_{0}\left(2\,I^{\prime}_{2}-2\,I^{\prime}_{4}+\,I^{\prime}_{6}\right)
\end{equation}

where $\,I^{\prime}_{n}\left(\,X;\,x_{0}\right)=\,I^{\prime}_{n}\left(0;10\right)$ for $\,n=2,4,6$.

\begin{equation}
 \,I^{\prime}_{2}\left(0;10\right)=0.623, \quad \,I^{\prime}_{4}\left(0;10\right)=0.123, \quad \,I^{\prime}_{6}\left(0;10\right)=0.029
\end{equation}

Hence, inserting these values in (\ref{EQ}) we obtain

\begin{eqnarray}
 &&\,E=\,E_{rest}+1.029\lambda_{0}\nonumber\\
&&=2.058+1.029\lambda_{0},
\end{eqnarray}

where 2.058 is the rest energy of the Q-ball with $\omega=\sqrt{3}$.
Using this equation to calculate the rest energy of the Q-ball on the top of a barrier or inside a hole, we found a complete agreement between the calculated and the numerical values.

Table 7  and table 8 show the calculated and numerical rest energies of the Q-ball on the top of a barrier and inside a hole respectively for various magnitude of the potential parameter,$\lambda_{0}$.

\begin{table}
\begin{center}
\begin{tabular}{ccc}\\ [3pt]
\hline
$\lambda_{0}$ & calculated $\,E_{B}$ & observed $\,E_{B}$   \\ [3pt]
\hline
0.1 &2.1609 & 2.1609  \\
 0.075 &2.1352  & 2.1352   \\
 0.05  & 2.1095 & 2.1095  \\ [3pt]
\hline
\end{tabular}
\end{center}
\caption{The calculated and observed rest energies of a Q-ball at the top of a barrier, $\,E_{B}$}
\end{table}

\begin{table}
\begin{center}
\begin{tabular}{ccc} \\[3pt]
\hline
$\lambda_{0}$ & calculated $\,E_{H}$ & observed $\,E_{H}$   \\ [3pt]
\hline
 -0.1 &1.955 & 1.955  \\
 -0.075 & 1.9808 &  1.9808 \\
 -0.05  & 2.0066 & 2.0066  \\ [3pt]
\hline
\end{tabular}
\end{center}
\caption{The calculated and observed rest energies of a Q-ball inside a hole, $\,E_{H}$}
\end{table}

 The critical velocity can be calculated using (114)
\begin{eqnarray}
&&\,u_{c}=\sqrt{1-\left(\frac{\,E_{rest}}{\,E}\right)^{2}}\nonumber\\
&&=\sqrt{1-\left(\frac{2.058}{2.058+1.029\lambda_{0}}\right)^{2}}.
\end{eqnarray}

For a barrier height of $\lambda_{0}=0.01$, the critical velocity, according to the above equation, for the Q-ball solution with $\omega=\sqrt{3}$ is $\sim 0.1$ which is the same as the numerical value \cite{AZ09}.

\section{Conclusion}
we analysed the dynamics of some the soliton-obstruction systems that we have studied via numerical simulations. We were able to demonstrate that our approximated models were successfully able to explain most of the observed behaviour of solitons scattering off potential obstructions. We have approached the dynamics of such systems in (1+1) dimensions by using collective coordinates, {\it ie} parameters of the theory. In the case of topological solitons, $\lambda\phi^{4}$ model and deformed Sine-Gordon model of class I potentials, the position of soliton is the only parameter that has been used to approximate their models. More than one parameter can be used to approximate such models but this will result in further analytical complications. In the case of Q-ball in (1+1) dimensions two parameters were at least needed to explain the dynamics namely: the position of the Q-ball and the phase which is latter identified as the angular frequency of the Q-ball. Our analytical work was able to explain the observed rest masses in barriers and holes and to produce trajectories that matches the ones produced by simulations to a great extent and to find out the forces between the solitons and potential obstructions. The forces and the potentials exerted on the solitons due to the obstructions were plotted for each model.  Further work can be done on building better approximated models which we will leave it for the future. 

\textbf{Acknowledgement}
 I would like to thank W.J. Zakrzewski for a helpful discussion.

\end{document}